\documentclass[12pt]{article}

\usepackage{amssymb,amsmath}
\usepackage{graphicx,epsf}

\makeatletter

\@addtoreset{equation}{section} \makeatother

\usepackage{graphicx}
\begin{document}

\begin{flushright}
LPT-ORSAY-04-63
\end{flushright}
\vskip 3pt

\begin{center}
{\Large \bf Brane cosmology with an anisotropic bulk} \vskip 1cm
A.~Fabbri$^{a,}$\footnote{{\tt Alessandro.Fabbri@bo.infn.it}},
D.~Langlois$^{b,c,}$\footnote{{\tt langlois@iap.fr}} ,
D.A.~Steer$^{d,c,}$\footnote{{\tt steer@th.u-psud.fr}} and
R.~Zegers$^{d,c,}$\footnote{{\tt robin@th.u-psud.fr}} \vskip 5pt
\vskip 0.7cm {\it a}) Dipartimento di Fisica dell'Universit\`a di
Bologna,\\ and INFN sezione di
Bologna, Via Irnerio 46, 40126 Bologna, Italy\\
and
\\
Departamento de F\'{\i}sica Te\'orica, Facultad de F\'{\i}sica, Universidad de
  Valencia, Burjassot-46100, Valencia, Spain.
\vskip 3pt
 {\it b}) Institut d'Astrophysique de Paris,
GR$\varepsilon$CO, FRE 2435-CNRS, 98bis boulevard Arago, 75014
Paris, France.
\vskip 3pt
{\it c}) F\'ed\'eration de recherche APC, Universit\'e Paris VII,\\
2 place Jussieu - 75251 Paris Cedex 05, France. \vskip 3pt {\it
d}) Laboratoire de Physique Th\'eorique\footnote{Unit\'e Mixte de
Recherche du CNRS (UMR 8627).}, B\^at. 210, Universit\'e
Paris XI, \\ 91405 Orsay Cedex, France.\\
\end{center}

\def\beq{\begin{equation}}
\def\eeq{\end{equation}}
\newcommand{\bea}{\begin{eqnarray}}
\newcommand{\eea}{\end{eqnarray}}

\def\B{{\bf B}}
\def\l{{\boldsymbol{\lambda}}}
\def\C{{\cal C}}
\def\A{{\cal A}}
\def\c{{c}}
\def\bb{{\boldsymbol{\beta}}}
\def\Bb{{\boldsymbol{\cal B}}}
\def\b{{\rm b}}

\newcommand{\be}{\begin{equation}}
\newcommand{\ee}{\end{equation}}
\newcommand{\ba}{\begin{eqnarray}}
\newcommand{\ea}{\end{eqnarray}}


\vskip 2.0cm

\begin{abstract}

In the context of brane cosmology, a scenario where our universe
is a $3+1$-dimensional surface (the ``brane'') embedded in a
five-dimensional spacetime (the ``bulk''), we study geometries
for which
the brane is anisotropic --- more specifically Bianchi I ---
though still homogeneous.
We first obtain explicit vacuum bulk solutions
 with anisotropic three-dimensional
spatial slices.  The bulk is assumed to be empty but endowed with
a negative cosmological constant. We then embed $Z_2$-symmetric
branes in the anisotropic spacetimes and discuss the constraints
on the brane energy-momentum tensor due to the five-dimensional
anisotropic geometry.  We show that if the bulk is static, an
anisotropic brane cannot support a perfect fluid.  However, we
find that for some of our bulk solutions it is possible to embed a
brane with a perfect fluid though  its energy density and pressure  are
completely determined by the bulk geometry.

\end{abstract}
\renewcommand{\thefootnote}{\arabic{footnote}} \setcounter{footnote}{0}

\newpage

\section{Introduction}

Since the advent of brane models with non-compact extra
dimensions, brane world cosmology has been studied in depth
(see \cite{Reviews1,Reviews2,Reviews3} for reviews).
The
typical setup is one of a homogeneous and isotropic brane
containing a perfect fluid, and the brane Friedmann equation can
be obtained in (at least) two different ways.  In the first
approach \cite{BDL1,BDL2}, coordinates are chosen relative to
the brane which is therefore at a fixed
position in the extra dimension. The bulk, on
the contrary is apparently
time dependent and this time dependence induces,
via the junction conditions, time dependence and hence cosmology
on the brane. In the second approach \cite{Kraus,Ida},
the bulk is static and the
brane moves along the extra dimension with this motion now being
responsible for the cosmology.  As was shown in \cite{Christ}, the
symmetries (homogeneity and isotropy) of the brane impose that the
static bulk is necessarily Sch-AdS.  Furthermore these two
approaches are completely equivalent via a coordinate
transformation \cite{Japs}, and Birkhoff's theorem applies
\cite{Christ}.

As in standard 4D cosmology, an obvious generalisation of such
models is to consider homogeneous but anisotropic brane worlds
\cite{MSS,Ruth1,Ruth2,Paul,Mak,Filippo,Frolov,CMMS,BH,CH,Halpern,Top,BD}.
Recall that in 4D, it is consistent with the Einstein equations to
study a universe containing a perfect fluid \cite{Ellis}. (We use
the standard definition of a perfect fluid for which there are no
anisotropic stresses or energy fluxes so that $T^{\mu}_{\nu} =
{\rm diag}(-\rho,P,P,P)$ with $\rho$ the energy density and $P$
the isotropic pressure.) In other words, geometric anisotropy is
consistent with matter isotropy in 4D. Thus it is possible, for
example, to study inflation due to a homogeneous scalar field and
ask whether the universe isotropises. As proved by Wald
\cite{Wald}, a positive cosmological constant will always
isotropise the universe providing that the perfect fluid matter
satisfies the weak and strong energy conditions. Observational
constraints on anisotropic universes are discussed in
\cite{const1,const2}.

The majority\footnote{The exceptions, \cite{Frolov,CMMS}, will be
discussed below.} of works on anisotropic
brane world universes use the {\it effective} four-dimensional
Einstein equations on the brane \cite{sms99}
\be
^{4}G_{a b} = \kappa_4^2 T_{ab} -{\cal
E}_{ab} + \kappa^4 S_{ab},
\label{G4}
\ee
where $^{4}G_{a b}$ is constructed from the induced metric on the
brane $\gamma_{ab}$, $S_{ab}$ is quadratic in the brane stress
energy tensor $T_{ab}$, and ${\cal E}_{ab}$ is the projection on
the brane of the bulk Weyl tensor. Although this equation looks
straightforward, it must be interpreted with care since the Weyl
term not only depends on brane quantities --- either geometry or
matter
--- but also on the bulk metric.  Thus the above brane equations
are not closed. More precisely, the crucial unknown quantity in
this effective approach is $\pi^*_{ab}$, the anisotropic component
of ${\cal E}_{ab}$, defined in the decomposition \cite{Maartens}
\be
-\frac{1}{\kappa_4^2} {\cal E}_{ab} = \rho^*\left( u_a u_b+
\frac{1}{3} h_{ab} \right) + \pi^*_{ab} + q^*_{(a} u_{b)}
\label{decomp}.
\ee
Here $u^\mu$ is a given 4-velocity, $h_{ab} = \gamma_{ab} + u_{a}
u_{b}$ and $\rho^*$, $\rho^*/3$ and $\pi^*_{ab}$ are the `dark'
energy density, pressure (${\cal E}_{ab}$ is traceless so it acts
as relativistic matter), and anisotropic stress respectively. The
dark momentum density $q^*_{a}$ vanishes for the Bianchi I branes
considered here. The sign of $\rho^*/3$ is also crucial for Wald's
theorem and brane isotropisation \cite{Filippo}.

In the literature, in order to  close the system of brane
equations (\ref{G4}), various additional conditions on
$\pi^*_{ab}$ have been proposed in a rather ad hoc way.
 Whether or not the universe isotropises, or indeed whether or not
inflation can take place at all depends on the choice made.
Similar comments hold for the isotropy of the initial singularity.

In order to have a fully consistent picture, one cannot avoid
specifying  the bulk geometry. This has non trivial implications
for the brane itself because, in brane world models, the Israel
junction conditions --- relating the extrinsic curvature to the
matter on the brane --- must be satisfied.  For instance, suppose
that the brane contains perfect fluid matter.  Then the junction
conditions will impose given relationships between the components
of the extrinsic curvature: are these relations, together with the
bulk Einstein equations, in fact compatible with an anisotropic
brane?  One of the initial motivations of this work was to try
and answer that question.

A first step was taken in \cite{CMMS}. There the authors considered a moving
brane in a static anisotropic bulk.  Having solved the bulk
Einstein equations, they then showed that the junction conditions
induce anisotropic stresses in the matter on the brane. Hence the
brane does not contain a perfect fluid. Furthermore, from the bulk
Einstein equations and the junction conditions, it follows that
this anisotropic stress can only vanish if the bulk is isotropic
and hence the brane is isotropic!  Thus, the conclusion is that
``geometric anisotropy enforces, via the extrinsic curvature and
the junction conditions, anisotropy of the matter fields''
\cite{CMMS}. Finally, note that the anisotropic stresses obtained
on the brane in \cite{CMMS} are {\it fixed} by the {\it bulk}
metric.  Thus, in general, they will not correspond to the
intrinsic properties of physical matter.

Another attempt was made in \cite{Frolov}.  For the choice of bulk
metric made there, it was shown that the brane can only be
anisotropic if it contains a constant tension, $\rho = P$!

In this paper we go one step further in the same direction.
We first start by  constructing new anisotropic solutions of the
5D bulk Einstein equations. The full five-dimensional Einstein
equations are much more difficult to solve when one assumes 3D
homogeneity {\it only}, rather than 3D homogeneity {\it and} isotropy
as in the first works on brane cosmology. However, with some
additional ans\"atze, we can find explicit geometries that
generalize the solutions previously obtained in the literature.

We then introduce a brane in the anisotropic bulk geometries and
study the constraints on the brane energy-momentum tensor.  In
particular, our aim is to see whether perfect fluid matter on the
brane is compatible with non-zero geometric anisotropy. We show
that a geometrically anisotropic brane moving in a {\it static
bulk} necessarily has a stress energy tensor which is {\it not}
that of a perfect fluid. Finally, we are able to construct a
configuration in which matter on the brane is purely described by
a perfect fluid, though its energy density and pressure as a
function of brane time are completely determined by the geometry.
As we will explain, this is not so surprising.

The paper is set up as follows. In section \ref{sec:bulk} we write
down explicitly the full system of Einstein equations which we aim
to solve. Then, in section \ref{sec:exact}, we solve them in two
particular cases: firstly we assume that the anisotropy depends
only on time and not on the extra-dimension; and secondly we
assume that the metric components are {\it separable} with respect
to time and to the extra-dimension. In section \ref{sec:brane} we
embed an anisotropic but homogeneous brane in the anisotropic bulk
solutions and discuss whether or not the brane matter can be a
perfect fluid. Conclusions are given in the final section.

\section{Bulk equations}
\label{sec:bulk}
We start from an ansatz for the bulk metric of the form
\beq
\label{metric} ds_{\rm bulk}^2=-e^{2A_0(t,w)}dt^2+\sum_{i=1}^3
e^{2A_i(t,w)}\left(dx^i\right)^2+dw^2,
\eeq
where the $x^i$ coordinates span the three ordinary spatial
dimensions and $w$ is the coordinate of the extra dimension.
Of course, one could start with a different choice of gauge, but
we have found the assumptions  $g_{ww}=1$ and $g_{w\mu}=0$
 convenient for integrating the bulk Einstein equations.

In the bulk, endowed with a (negative) cosmological constant $\Lambda\equiv
-6/\ell^2$, the five-dimensional Einstein's equations simply read
\beq
R_{AB}=\frac{2}{3}\Lambda g_{AB} \qquad \qquad (A=0,\ldots,4).
\eeq
Inserting the metric ansatz (\ref{metric}) into Einstein's
equations yields the following system of equations:
\bea
\label{Rtt} A_0''+A_0'\sum_{\mu=0}^3
A_\mu'+e^{-2A_0}\left[-\sum_{i=1}^3\ddot A_i
-\sum_{i=1}^3\dot A_i^2+\dot A_0\sum_{i=1}^3\dot A_i\right]&=&{\frac{4}{\ell^2}}\\
 \label{Rii} A_i''+A_i'\sum_{\mu=0}^3
A_\mu'+e^{-2A_0}\left[-\ddot A_i
-\dot A_i\sum_{j=1}^3\dot A_j+\dot A_0\dot A_i\right]&=&{4\over \ell^2}\\
\label{Rww}
\sum_{\mu=0}^3 A_\mu''+\sum_{\mu=0}^3 {A_\mu'}^2&=&{4\over \ell^2}\\
\label{Rtw} \sum_{i=1}^3{\dot A_i}'-A_0'\sum_{i=1}^3\dot
A_i+\sum_{i=1}^3\dot A_i A_i'&=&0,
\eea
where a dash denotes a derivative with respect to $w$, and a dot
one with respect to $t$.  Note that middle of the alphabet latin
indices label the three ordinary spatial directions in the bulk,
whilst greek indices also include the time component. In order to
solve these equations, it is convenient to introduce the average
scale factor $s \equiv (\exp(\sum_i A_i))^{1/3}$ of the ordinary
spatial directions $x^i$, and define
\beq \label{aver}
s \equiv e^{A} , \qquad A = {1\over 3}\sum_{i=1}^3 A_i.
\eeq
Furthermore we denote the deviation from isotropy by a vector
${\bf B}$ with components
\beq \label{bi}
B_i\equiv A_i-A,
\eeq
 so that
\beq
\label{sum} \sum_{i=1}^3 B_i=0.
\eeq
Thus $\B'$ and $\dot{\B}$ quantify the spatial and temporal
shear of the metric
(\ref{metric}).

The Einstein equations can now be reexpressed in terms of the
isotropic and anisotropic quantities we have just introduced.
Eq.~(\ref{Rtt}) becomes
\beq
\label{Rtt2}
A_0''+ A_0'\left( A_0'+3A'\right)+e^{-2A_0}\left[-3\ddot A -3\dot
A^2-\dot{\bf B}^2+3\dot A_0\dot A\right]={4\over \ell^2},
\eeq
while Eq.~(\ref{Rii}) can be decomposed into an ``isotropic'' part, obtained
by averaging over the index $i$,
\beq
\label{Rii_iso}
A''+A'\left( A_0'+3A'\right)+e^{-2A_0}\left[-\ddot A
-3\dot A^2+\dot A_0\dot A\right]={4\over \ell^2},
\eeq
and an ``anisotropic'' part that  can  be written as
\beq
\label{Rii_ani2}  \left[{\bf B}' e^{3A + A_0  }\right]' =
\left[\dot{\bf B} e^{3A - A_0}\right] ^{\dot{}}
\eeq
Finally, equations (\ref{Rww}) and (\ref{Rtw}) give respectively
\beq
\label{Rww2} A_0''+3A''+{A_0'}^2+3{A'}^2+{{\bf B}'}^2 ={4\over
\ell^2}
\eeq
and
\beq
\label{Rtw2} {\dot A}'+\left(A'-A_0'\right)\dot A+{1\over
3}\dot{\bf B}\cdot {\bf B}'=0.
\eeq
This rewriting of the equations easily allows us to see how
anisotropy modifies the usual equations for the bulk: setting the
anisotropic parts ${\bf B}$ to zero immediately yields the same
system of equations as in isotropic brane cosmology \cite{BDL1,BDL2}.

It is thus useful to try to generalise
the integration of Einstein's equations in the context of isotropic
brane cosmology to the present anisotropic case.
A combination of (\ref{Rtt2}), (\ref{Rii_iso}) and (\ref{Rww2}), in fact proportional
to the component $G_0^0$ of the Einstein tensor, gives
\beq
\label{Gtt}
A''+2{A'}^2+{1\over 6}{\B'}^2
+e^{-2A_0}\left(-\dot A^2+{1\over 6}\, {\dot\B}^2\right)=
{2\over \ell^2}.
\eeq
Another combination of the same equations,
corresponding to the component $G_w^w$ of Einstein's
tensor, gives
\beq
\label{Gww}
{A'}^2+A_0'A'-{1\over 6}{\B'}^2+e^{-2A_0}\left(-\ddot A-2\dot A^2+\dot A_0
\dot A- {1\over 6}{\dot\B}^2\right)=
{2\over \ell^2}.
\eeq
{ Following \cite{BDL2} let us now introduce the quantity}
\beq
F\equiv e^{4A}\left({A'}^2-e^{-2A_0}\dot A^2\right).
\eeq
Then on using (\ref{Rtw}), one can rewrite (\ref{Gtt}), after
multiplication by $A'$, as
\beq
\label{Fprim}
F'e^{-4A}+{1\over 3}{\B'}^2A'+{1\over 3}e^{-2A_0}\left({\dot\B}^2A'-2\dot A
\, \dot\B.\B'\right)={4\over \ell^2}A'.
\eeq
Similarly, (\ref{Gww}) yields
\beq
\label{Fdot}
\dot Fe^{-4A}-{1\over 3}{\B'}^2\dot A
-{1\over 3}e^{-2A_0}{\dot\B}^2\dot A
+{2\over 3} A'\,
\dot\B.\B'={4\over \ell^2}\dot A.
\eeq
Note that when $\B=0$, one recognises the results of isotropic
brane cosmology \cite{BDL2} in which case the two above equations
can readily be integrated. In the anisotropic case, these
equations are much more difficult and, in order to integrate them,
we will have to resort to simplifying assumptions as discussed in
the next section.

\section{Exact bulk solutions}
\label{sec:exact}
In this section we make, in turn, two  particular assumptions
about the metric: these will enable us to integrate Einstein's
equation explicitly.  First we assume that the shear does not
depend on the extra-dimension, namely that $\B'=0$. Then we
consider the situation in which the metric is separable, that is all
the metric coefficients can be expressed as the product of a
function of $t$ with a function of $w$.

\subsection{Case $\B'=0$}
\label{sec:case}
When the anisotropic parts are only time-dependent, i.e. $\B'=0$,
then the equations established in the previous section simplify
greatly. This situation is similar to that studied in \cite{CR} in
the case of Weyl metrics.

With $\B'=0$, Eq.~(\ref{Rii_ani2})
can be integrated to give
\beq
\dot {\bf B} =\l(w) e^{A_0-3A} \label{intB}
\eeq
where $\l(w)$ is an arbitrary function of $w$. On substituting
into (\ref{Fprim}) and (\ref{Fdot}) we obtain
\begin{equation}
\label{Fprim2} F'e^{-4A}+{1\over 3}{\l}^2 e^{-6A}A'={4\over
\ell^2}A' \end{equation} and
\beq
\label{Fdot2}
\dot Fe^{-4A}
-{1\over 3}\l^2 e^{-6A} \dot A
={4\over \ell^2}\dot A.
\eeq
Since $\l$ is a function of $w$ only, equation (\ref{Fdot2})
can be integrated in time to yield
\beq
\label{Fint} F+{1\over 6}\l^2 e^{-2A}={1\over \ell^2}e^{4A}+
\C(w),
\eeq
where $\C(w)$ is an arbitrary function of $w$. Substitution of
this first integral in (\ref{Fprim2}) gives a consistency relation
between the integration functions $\l(w)$ and $\C(w)$,
\beq
{1\over 3}e^{-2A}\left(\l.\l'-2 A'\l^2\right)=\C'
\eeq
or alternatively, using the relation $\l=\dot\B e^{-A_0+3A}$,
\beq
\label{consteq}
 e^{-2A}\left(A'-A_0'\right)= 3
\frac{\C'(w)}{\l^2(w)}.
\eeq
Since $\l$ and $\C$ are only $w$-dependent while $A$ and $A_0$ in
general are time-dependent, a simple way to satisfy this
constraint is if both sides of (\ref{consteq}) vanish. { The more
general case does not seem to yield more non-trivial solutions}.

Assuming, therefore, that  $\C$ is a constant and $A'=A_0'$,
equation (\ref{Rtw2}) implies $\dot A'=0$ and therefore
\beq
A=\alpha(t)+\A(w), \quad A_0=\alpha_0(t)+\A(w).
\eeq
This means that the  metric is  of the form
\begin{eqnarray} \label{fgauge}
ds^2&=&e^{2\A(w)}
\left[-e^{2\alpha_0(t)}dt^2+\sum_i e^{2\alpha_i(t)}{dx^i}^2\right]+dw^2\\
&\equiv & a^2(w) h_{\mu\nu}(t) dx^\mu dx^\nu+ dw^2,
\end{eqnarray}
which shows that the dependence on the extra-dimension reduces to a single
warping factor. In this sense, the anisotropy
will be the same on all slices $w=$ const, simply
rescaled by this warping factor.

We now solve explicitly for the warping factor $a(w)=\exp\A(w)$.
From (\ref{Rww}), or equivalently (\ref{Rww2}), it
 satisfies the equation
\beq
a''={1\over \ell^2} a,
\eeq
so that
\beq {a'}^2={1\over \ell^2}a^2+\c,
\eeq
where $\c$ is an integration constant. Depending on the sign of
$\c$ there are three solutions
\begin{eqnarray}
\label{asoln}
 a(w) = \left\{ \begin{array}{ccc}
\sqrt{\c}\, \ell \sinh(w/\ell) &\quad c>0, \\
\sqrt{-\c}\, \ell \cosh(w/\ell) &\quad c<0, \\
\exp(w/\ell) &\quad c=0, \\
\end{array} \right.
\end{eqnarray}
where we have absorbed another integration constant in a redefinition
(by translation) of $w$, and $w$ is also defined up to a sign.
The other components of Einstein's equations can be rewritten in the
form
\beq
{}^{(5)}R_{\mu\nu}={}^{(4)}R_{\mu\nu}- \left(a a''+3{a'}^2\right)h_{\mu\nu}
=-{4\over \ell^2} a^2 h_{\mu\nu}.
\eeq
where ${}^{(4)}R_{\mu\nu}$ is the Ricci tensor for the
four-dimensional metric $h_{\mu\nu}$. This implies
\beq \label{eqs4d}
{}^{(4)}R_{\mu\nu}=(3\c) h_{\mu\nu}\  \eeq so that $h_{\mu \nu}$
satisfies the 4D Einstein equations with cosmological constant
$6c$.

We now turn  to the time dependent functions $\alpha_i(t)$ in
(\ref{fgauge}). Note that one can set $\alpha_0=0$ by a
redefinition of
 the time coordinate, and this will be assumed below. For arbitrary $c$, let
\beq
\alpha_i (t)= \alpha(t) + \beta_i (t) \ , \eeq where
$\alpha(t)=\sum_i \alpha_i(t)/3$ and $\sum_i
\beta_i(t)=0$ in analogy with equations
(\ref{aver}-\ref{sum}).
Equation (\ref{intB}) can now be written in the form
\be
\dot{\beta}_i = b_i e^{-3\alpha(t)} \label{tre}
\ee
where the integration constants $b_i$ satisfy $\sum_{i}b_i=0$. The
remaining Einstein equations reduce to
\be
\label{uno} 3 \ddot \alpha + 3 \dot\alpha^2 + {\bf b}^2
e^{-6\alpha} = 3\c \qquad {\rm and} \qquad \ddot\alpha +
3\dot\alpha^2 = 3\c
\ee
as given by (\ref{eqs4d}),
so that
\beq
\dot{\alpha}^2 = c + {\bf b}^2 e^{-6\alpha}/6.
\eeq
On integrating it follows that, for ${\bf b}^2\neq 0$,
\begin{eqnarray}
\label{alphas} e^{3\alpha(t)}= \left\{ \begin{array}{ccc}
\sqrt\frac{{\bf b}^2}{6{\c}} \sinh (3\sqrt{\c} t)  &\quad c>0, \\
\sqrt\frac{{\bf b}^2}{6(-\c)} \sin (3\sqrt{-\c} t) &\quad c<0, \\
\sqrt\frac{{\bf b}^2}{6}
(3t) &\quad c=0, \\
\end{array} \right.
\end{eqnarray}
and the $\beta_i$ are given by (\ref{tre})
\begin{eqnarray}
\label{betas}
\beta_i= \frac{\tilde b_i}{3}
 \left\{ \begin{array}{ccc}
\ln\tanh (\frac{3}{2}\sqrt{\c} t) &\quad c>0, \\
 \ln\tan
(\frac{3}{2}\sqrt{-\c} t) &\quad c<0,\\  \ln t &\quad c=0.\\
\end{array} \right.
\end{eqnarray}
Here we have introduced a renormalised $\tilde{b}_i$, related to
$b_i$ by
\beq
{\tilde b_i}\equiv \sqrt{\frac{6}{{\bf b}^2}}\, b_i,
\eeq
so that
\beq
\sum_i {\tilde b_i}=0, \qquad \qquad \sum_i {\tilde b_i}^2=6.
\label{anto}
\eeq

Putting together the dependence on the extra-dimension
(\ref{asoln}) and the time dependence (\ref{alphas}) and
(\ref{betas}), the full metric in the simplest case when the
effective cosmological constant on the brane $c$ vanishes is given by
\beq
\label{warped_Kasner}
ds^2=e^{2w/\ell}\left[-dt^2+\sum_i t^{2p_i}
\left(dx^i\right)^{2}\right] + dw^2
\eeq
with
\be
 \sum_i
p_i=1,\qquad \sum_i p_i^2=1.
\ee
We have replaced the ${\tilde b}_i$ by $p_i=({1+{\tilde b}_i})/{3}$,
so as to  recognise a warped version of the usual 4D Kasner metric
\cite{Frolov}.

When the effective cosmological constant is positive, $c>0$, the full metric
can be written, after some appropriate rescalings, in the form
\beq
ds^2=\sinh^2(w/\ell)\left[-dt^2+\sum_i\sinh^{2/3}(3t/\ell)
\left(\tanh\left({3t\over 2\ell}\right)\right)^{2\tilde
b_i/3}\left(dx^i\right)^{2} \right] + dw^2.
\eeq
Finally, the case of a negative cosmological constant
is simply obtained by converting the hyperbolic functions into
trigonometric ones.

\subsection{Separable solution}
\label{sec:sep}

Let us now go back to Einstein's equations in the form
(\ref{Rtt}-\ref{Rtw}) and look for separable solutions
\beq
\label{sep}
 A_\mu(t,w)=\alpha_\mu(t)+\A_\mu(w).
\eeq
This time we do not assume that the spatial variation of the
anisotropy, ${\bf B}'$, vanishes and the solutions of the previous
subsection, which eventually turned out to be separable, are a
priori only a subclass of the separable solutions.

In the equations (\ref{Rtt}) and (\ref{Rii}), we further assume
that the brackets involving time derivatives separately
vanish.\footnote{A priori, in the gauge $\alpha_0=0$, one could
consider the more general case where the brackets are equal to
four different constants.  However, it can then be shown that
these constants are necessarily equal, and that if the resulting
single constant does not vanish then one recovers the situation
considered in the previous section.} Leaving apart for now the
mixed equation (\ref{Rtw}) the equations involving {\it spatial
derivatives} reduce to  the following system:
\bea
&& \label{wRaa}
\A_\mu''+\A_\mu'\sum_{\nu=0}^3 \A_\nu'={4\over \ell^2},\\
&&\label{wRww} \sum_{\mu=0}^3 \A_\mu''+\sum_{\mu=0}^3
{\A_\mu'}^2={4\over \ell^2}.
\eea
Although here we allow for a time dependence, these equations are
exactly the same as those obtained from Einstein's equations when
assuming a {\it static} ansatz, which is the situation studied in 
\cite{CMMS}. We
can thus follow the method of \cite{CMMS} to integrate the {\it
spatial} dependence  of the metric. First introduce the quantity
\begin{equation}
u(w) = \exp\left(\sum_{\mu=0}^3 \A_\mu \right)\,. \label{u}
\end{equation}
Multiplying Eq.~(\ref{wRaa}) by $u$ and summing over $\mu$ gives
\begin{equation}
  u''-{16\over \ell^2} u =  0\,,
\end{equation}
which admits the first integral,
\begin{equation}
  u'^2 - {16\over \ell^2} \left( u^2 + \gamma^2 \right) =  0
\label{eq:integral}
\end{equation}
where we have written the integration constant as
$-16\gamma^2/\ell^2$ for later convenience (we will also show that
$\gamma^2\geq 0$). One can then integrate (\ref{wRaa}) to obtain
\begin{equation}
\A'_\mu = \frac{1}{4}\frac{u'}{u}+ \frac{4 \gamma}{ \, \ell}
\frac{q_\mu}{u}\, , \label{eq:A'}
\end{equation}
where the $q_{\mu}$ are integration constants.

When $\gamma=0$, equations (\ref{eq:integral}) and (\ref{eq:A'})
yield
\begin{equation}
\A_\mu(w) = w/\ell \,,
\end{equation}
where the  integration constants have been absorbed in a rescaling
of the coordinates.  This corresponds to the particular case $c=0$
obtained in  the previous subsection.

For arbitrary $\gamma$, the constants $q_\mu$ are constrained by
Eqs.~(\ref{u}), (\ref{wRww}) and (\ref{wRaa}):
\beq
\label{qcon} \sum_\mu q_\mu=0, \qquad \sum_\mu q_\mu^2={3\over 4}
\, ,
\eeq
implying $| q_\mu| \leq {3}/{4}$.
The second equality justifies the
positivity of $\gamma^2$.
Equation (\ref{eq:integral}) then yields
\begin{equation}
u(w) = \gamma \sinh \left({4w/ \ell} \right) \,,
\label{eq:u_sol}
\end{equation}
where we have absorbed an integration constant in $w$. Note
that replacing $w$ by $-w$ also gives a valid solution. Thus
Eq.~(\ref{eq:A'}) gives
\begin{equation}
\A_\mu(w) =  \frac{1}{4}\ln \left|u(w)\right| +  \frac{4
\gamma}{\ell} q_\mu v(w)+ {\rm const}\,,
\end{equation}
where $v'=1/u $, so that
\begin{equation}
v(w) =  \frac{\ell}{4 \gamma} \ln \tanh(2|w|/\ell)
\, .
\end{equation}

Finally, we thus get
\beq
e^{2\A_\mu} = N^2_\mu \sinh^{1/2}(4|w|/\ell)
\tanh^{2q_\mu} (2|w|/\ell)
 \,, \label{met_coef}
\eeq
where  $N_\mu$ are integration constants
 which must satisfy the constraint
\begin{equation}
\prod_{\mu=0}^3 N^2_\mu = \gamma^2 \,,
\end{equation}
following from Eqs.~(\ref{u}) and (\ref{eq:u_sol}).
Note that replacing
$w$ by $-w$ in (\ref{eq:u_sol}) would
yield (\ref{met_coef}) with $-q_\mu$ in place
of $q_\mu$.

{ As mentioned above,
when the metric is assumed to be {\it static}, the full Einstein
equations reduce to the system (\ref{wRaa}-\ref{wRww}) and
the above results are enough to determine  the full metric.
Note that as discussed in \cite{CMMS}, the case
$(q_0,q_i)=(-3/4,1/4)$ corresponds to Sch-AdS$_5$.}

Here we assume that the metric is {\it not static} and thus the
other Einstein equations must be integrated in order to determine
 the time dependence of the metric. Given the separable ansatz
(\ref{sep}), it is easy to see that the time-dependent part of the
metric components, imposing the gauge $\alpha_0=0$, must satisfy
exactly the same equations as in section \ref{sec:case} with
$c=0$. The solutions are given in (\ref{alphas}) and
(\ref{betas}), with $c=0$. However, there is also  a further
constraint coming from Eq.~(\ref{Rtw}) which relates time and
space derivatives:
\beq
\label{ant} \sum_i q_i({\tilde b}_i+4)=0.
\eeq
Note that when all the $q_i$'s are equal but non-zero, then
condition (\ref{ant}) is incompatible with (\ref{anto}), which
indicates that Sch-AdS$_5$ does not have a simple time dependent
extension.
  In
general, however, constraints (\ref{anto}), (\ref{qcon}) and (\ref{ant})
  can all be satisfied simultaneously
leading to time-dependent anisotropic solutions.

To summarize, the bulk solutions we have obtained are described by
a metric of the form
\beq
\label{metric_separable} ds^2=\sinh^{1/2}(4w/\ell)\left[- \tanh^{2
q_0}\left(2w/\ell\right) dt^2 +\sum_i \tanh^{2
q_i}\left(2w/\ell\right)t^{2p_i}\left(dx^i\right)^2\right] +dw^2,
\eeq
where the seven  coefficients $q_\mu$ and $p_i$ must satisfy the
following five constraints
\beq
\label{constraints}
\sum_\mu q_\mu=0, \quad \sum_\mu q_\mu^2={3\over 4}, \quad
\sum_i p_i=1 ,
\quad \sum_i p_i^2=1, \quad \sum_i q_i\left(p_i+1\right)=0.
\eeq
One can solve explicitly this above system of constraints
and, in appendix A, we give
the general solution for the coefficients in terms of two
parameters.

Interestingly, this metric ``mixes'' the five-dimensional static
solution of \cite{CMMS} and the well-known four-dimensional Kasner
solution. In this sense, we have found a much more sophisticated
``warped'' version of Kasner than in the previous subsection,
because there are now four different warp factors, along the time
and the three ordinary spatial directions.

As a final remark in this section, note that the above metric can
be rewritten in the form
\beq
\label{metric_z}
ds^2=\sqrt{2z\over 1-z^2}\left[- z^{2q_0}dt^2
+\sum_i z^{2 q_i}t^{2p_i}\left(dx^i\right)^2\right]
+{\ell^2\over 4(1-z^2)^2}dz^2,
\eeq
where $z=\tanh(2w/\ell)$. In appendix B, we have used this form to
compute the square of the Weyl tensor, which is gauge-invariant
and  thus  useful to analyse the physical singularities of the
metric.

\section{The brane}
\label{sec:brane}
So far we have obtained explicit vacuum solutions for the bulk
with a negative cosmological constant. In this section, we
consider an infinitely thin brane
 embedded in anisotropic bulk geometries,
and for simplicity study configurations with $Z_2$ symmetry about
the brane. We start by establishing the general junction
conditions, and then discuss the possibility of embedding a brane
with only a perfect fluid as matter.

\subsection{The embedding and junction conditions}

Denoting  the energy-momentum tensor of the brane matter  as
$T^a_b$, the Israel junction conditions, in the
case of $Z_2$ symmetry,
are given by
\be
K^{a}_b = - \frac{\kappa^2}{2} \left(T^{a}_{b} -
\frac{T}{3} \delta^{a}_{b}\right) \label{junction}
\ee
where $K^{a}_b$ is the extrinsic curvature on one side of the brane.
 We use lowercase
latin letters to denote the indices of the intrinsic coordinates
on the brane. In general, the geometry of the brane can be defined
by its embedding in the bulk space time, i.e. $X^A=X^A(x^a)$ where
the $x^a$ are the intrinsic brane coordinates. The extrinsic
curvature is then given by
\be
K_{ab} \equiv X^{A}_{a}X^B_{b} D_{A}n_B = \frac{1}{2}
\left[ g_{AB} (X^{A}_{a} \partial_b n^B + X^{A}_{b}
\partial_a n^B ) + X^A_a X^B_b n^C g_{AB,C} \right],
\label{symmK}
\ee
where $D_A$ is the covariant derivative associated with the bulk
metric $g_{AB}$, $n^A$ is the unit vector normal to the brane, and
$X^{A}_{a}\equiv \partial X^A/\partial x^a$.

If matter on the brane is a perfect fluid, 
\be
T^a_b={\rm
diag}(-\rho, P, P,P),\label{pf}
\ee
then the junction conditions
(\ref{junction}) imply that
 the spatial components of $K^{a}_b$ must be equal
\be
K^{1}_1 = K^2_2 = K^3_3 = - \frac{\kappa^2}{6} \rho,
\label{con-perf}
\ee
whilst the off diagonal components vanish. These conditions must
be satisfied at the brane position  for all times.

Let us now introduce  the bulk metric in the form
\beq
ds_{\rm bulk}^2 = -e^{2A_0(t,w)} dt^2 + \sum_i e^{2A_i(t,w)}
(dx^i)^2 + e^{2A_4(t,w)}dw^2, \label{ansatz}
\eeq
where, in order to be more general, we have kept $g_{ww}$  free.
To obtain a anisotropic but {\it homogeneous} brane,
 we  consider the  embedding
\be
X^A = (t_\b(\tau),{\bf x}, w_\b(\tau)) \label{embed},
\ee
where the subscript ``$\b$'' stands for brane. The coordinates
$\tau$ and ${\bf x}$ are the intrinsic brane coordinates. It is
always possible to choose the time parameter $\tau$ to be the
proper time by imposing the condition
\be
 e^{2A_0}\left( \frac{dt_\b}{d\tau}\right)^2   - e^{2A_4}\left(
\frac{dw_\b}{d\tau}\right)^2   =1 \label{proper}.
\ee
As a consequence,  the
induced metric on the brane is simply given by
\be
ds_{\rm brane}^2 =-d\tau^2 + \sum_i e^{2A_i(\tau)} (dx^i)^2
\ee
where $A_i(\tau) = A_i(t_\b(\tau), w_\b(\tau))$.

The shear in  the bulk also induces a shear in the brane geometry,
which can be expressed as
 $\sigma^{a}_{b} = {\rm diag}(0,\sigma^i)$, where
\be
\sigma_i \equiv  \frac{d}{d\tau}B_i(t_\b(\tau), w_\b(\tau)) =
\left. \dot{B_i} \right|_X \frac{dt_b}{d\tau} + \left. B_i'
\right|_X \frac{dw_b}{d\tau}. \label{bshear}
\ee
The brane is isotropic when the shear vanishes.  Similarly, the
average scale factor on the brane, $H$, is given by
\be
H \equiv  \frac{dA}{d\tau} = \left. \dot{A} \right|_X
\frac{dt_b}{d\tau} + \left. A' \right|_X \frac{dw_b}{d\tau}.
\label{Haver}
\ee

For the embedding (\ref{embed}) and metric (\ref{ansatz}) the
normal to the brane, $n^A$, is
\be
n^A = -\left(e^{-A_0+A_4} \dot w_\b, {\bf 0}, e^{A_0-A_4} \dot t_\b
\right)= -\left(e^{-A_0+A_4} \dot w_\b, {\bf 0},
e^{-A_4}\sqrt{1+e^{2 A_4} \dot w_\b^2} \right),
\ee
where a dot on $t_\b$ and $w_\b$ denotes a derivation with respect
to the {\it proper time} (and not the bulk time as for the bulk
quantities) and where the second equality follows from the
condition (\ref{proper}).  Here we have chosen the normal with a negative
$w$-component, meaning that we keep the bulk space-time with $w<w_b$.
For our particular bulk solutions, this condition will lead to $\rho>0$.
Using the definition (\ref{symmK}), one
can then compute the spatial components of the extrinsic curvature
tensor
\be
K^{i}_0=0, \qquad K^{i}_{j}=0 \; \; (i \neq j), \qquad K^i_{i} =
\frac{1}{2g_{ii}}\left( n^t \dot{g}_{ii} + n^w g'_{ii}  \right) =
n^t \dot{A}_i + n^w A_i'. \label{K}
\ee
The time/time component of the junction condition leads to the
usual energy conservation equation on the brane, $d\rho/d\tau + 3
H(\rho+P)=0$, where $H$ is the average Hubble parameter given in
(\ref{Haver}). This is the familiar result in the case of an empty
bulk and follows from the Gauss equation and (\ref{junction}).

{
The junction conditions (\ref{junction}) together with the components
of the extrinsic curvature tensor (\ref{K}) imply that the brane
energy-momentum tensor is necessarily of the form
\beq
T^a_b={\rm diag}\left(-\rho, P_1, P_2, P_3 \right),
\eeq
i.e.~all the off-diagonal terms are necessarily zero. However, notice that
this stress tensor is {\it generally anisotropic} with the pressure
depending a priori on the direction, {\it unlike the perfect fluid
form (\ref{pf})}. Let $P$ denote the isotropic pressure and
decompose
\be
P_i = P + \pi_i \label{anisostress}
\ee
where the anisotropic stresses $\pi_i$ satisfy $\sum_i \pi_i = 0$.
It is then instructive to decompose the spatial components of the
junction conditions (\ref{junction}) into an isotropic part,
\be
e^{-A_0+A_4}\dot w_b  \left. \dot{A} \right|_X + e^{-A_4}
\sqrt{1+e^{2A_4}\dot w_b^2 } \, \left. {A'} \right|_X =
\frac{\kappa^2}{6} \rho \label{Cond1},
\ee
and an anisotropic part,
\be
e^{-A_0+A_4}\dot w_b \left. \dot{B}_i \right|_X +e^{-A_4}
\sqrt{1+e^{2A_4}\dot w_b^2}\; \left. {B'}_i \right|_X =
\frac{\kappa^2}{2}\pi_i. \label{Cond2}
\ee
This last relation tells us that, in general, an anisotropic bulk
geometry implies the existence of an anisotropic stress on the
brane.

\subsection{Anisotropic brane with a perfect fluid?}
We can now study the  question of whether it is possible to find
a bulk geometry and a brane trajectory such that
(\ref{Cond2}) is satisfied with $\pi_i=0$,
\be
e^{-A_0+A_4}\dot w_b \left. \dot{B}_i \right|_X +e^{-A_4}
\sqrt{1+e^{2A_4}\dot w_b^2}\; \left. {B'}_i \right|_X  =0,
\label{Cond3}
\ee
that is, for a {\it perfect fluid}. In order to do this, we consider
in turn the bulk geometries constructed previously, as well as the
case of a static bulk geometry.

\subsubsection{A no-go condition: static bulk and a moving brane}
First consider  a static anisotropic bulk, described by the metric
\be
ds_{\rm bulk}^2 = -e^{2A_0(w)} dt^2 + \sum_i e^{2A_i(w)} (dx^i)^2
+ e^{2A_4(w)}dw^2.  \label{ansatz2}
\ee
Since we are interested in cosmology, the brane is necessarily
moving in this bulk, $\dot w_\b \neq 0$. As the bulk is static,
$\dot{\bf B}=0$, and condition (\ref{Cond3}) therefore reduces to
\be
\label{conn} \left. {{\bf B}'}  \right|_X = 0.
\ee
However, from (\ref{bshear}), this implies that
 the shear on the brane also vanishes and the
brane is  isotropic.

Thus  we conclude that it is not possible to have an anisotropic
moving brane containing a perfect fluid in a static background of
the form given in (\ref{ansatz2}). Notice that in order to reach
this conclusion it was not necessary to solve the bulk Einstein
equations: a moving brane containing a perfect fluid in a static
bulk is necessarily geometrically isotropic, and the bulk is
therefore Sch-AdS$_5$ \cite{Christ}.

Conversely, if one embeds a moving brane into such a static
anisotropic background then the brane  stress energy tensor is
not that of a perfect fluid: from the junction conditions, the
stress energy tensor picks up a bulk-dependent
 anisotropic stress as in \cite{CMMS}.  In other words,
the matter on the brane is fixed by the bulk geometry with the
anisotropic stress on the  brane, $\pi_{i}
\propto \left. B'_{i} \right|_X$.

\subsubsection{Bulk with ${\bf B}'=0$}\label{sec:braneB}

We now consider a bulk characterized by
 ${\bf B}'=0$, as in section \ref{sec:case}.
  Condition (\ref{Cond3}) then reduces to
\be
\left.\dot{w_\b}\,
\dot{\bf B}\right|_X={\bf 0}
\ee
so that there can only be a non-zero shear on the brane (see
(\ref{bshear})) if the brane is at a fixed position in the extra
dimension,
\beq
w_\b={\rm  constant}.
\eeq
As discussed in section \ref{sec:case}, when ${\bf B}'=0$ the bulk
Einstein equations impose that the bulk metric is separable (see
(3.7)).
  This implies that
condition (\ref{Cond1}) reduces to the time independent expression
\be
\rho =  A'(w_b) \frac{6}{\kappa^2}, \label{co}
\ee
because $A'$ depends only on $w$ and not $t$. The brane can
therefore contain a perfect fluid but it must be a constant
tension given by (\ref{co}).  The geometry on the brane is
anisotropic --- on choosing $w_b$ such that $a^2(w_b)=1$,
\be
ds_{\rm brane}^2 =-d\tau^2 + e^{2\alpha(\tau)} \sum_i
e^{2\beta_i(\tau)} (dx^i)^2
\ee
where $\alpha(\tau)$ and $\beta_i(\tau)$ are given in
(\ref{alphas}) and (\ref{betas}). Finally, note that for these
solutions the projected bulk anisotropic stress on the brane,
$\pi^*_{ab}$, vanishes since it is proportional to $\left. {\bf
B}'' \right|_{w_b}$.

\subsubsection{Separable bulk solution}
We now assume that the bulk geometry is given by the metric
(\ref{metric_separable}), that is $A_4=0$ and
\bea
e^{A_0}&=&\sinh^{1/4}(4w/\ell)\tanh^{q_0}\left(2w/\ell\right),\\
e^A&=&\sinh^{1/4}(4w/\ell)\tanh^{-q_0/3} \left(2w/\ell\right)t^{1/3},\\
e^{B_i}&=& \tanh^{q_i+q_0/3}\left(2w/\ell\right) t^{p_i-1/3}.
\eea
Expression (\ref{Cond3}) summarises three equations, one for each
value of $i$ which, for this bulk metric, yield
\beq
\label{Bast1} \frac{e^{-A_0} \, \dot w_\b }{\sqrt{1+\dot w_\b^2}}=
-\left.  {B_i'  \over \dot B_i  } \right|_X = -{4\over
\ell}\left(\frac{q_i + q_0 / 3}{p_i-1/3 }\right) \, \frac{
t_\b}{\sinh(4 w_\b/\ell)}.
\eeq
Clearly the three relations in (\ref{Bast1}) are only compatible
if the right-hand side is independent of $i$. Thus the
coefficients $p_i$ and $q_i$ must satisfy the relation
\beq
\label{Bast2} q_i+\frac{q_0}{3}=k \left(p_i-{1\over 3}\right) \,
\ee
where $k$ is a constant. Combining this with the constraints on
$p_i$ and $q_i$ given in (\ref{constraints}), or using the results
of appendix A, implies
\beq
q_0 = \pm \frac{ \sqrt{3}}{4}
\eeq
and thus $k=\pm \sqrt{3}/2$. Remarkably, therefore, the brane {\it
can support perfect fluid type matter} for  $q_0=\pm \frac{
\sqrt{3}}{4}$. As shown explicitly in appendix A, it is possible
to find sets of coefficients $(p_i,q_i)$ that satisfy all the
constraints plus the additional condition $q_0= \pm \sqrt{3}/{4}$.
All the possible solutions are expressed in terms of a single
parameter.

Of course, the brane cannot support  {\it any} perfect fluid, in
contrast with the isotropic case. The reason is that the
anisotropic junction condition (\ref{Cond3}) determines the
trajectory of the brane. Indeed, on substituting  $k=\pm
\sqrt{3}/2$, this relation now becomes
\beq
\label{wdot}
{\dot w_\b\over \sqrt{1+\dot w_\b^2}}=\mp {2\sqrt{3}\over\ell}t_b
{e^{A_0}\over \sinh(4 w_b/\ell)} \equiv f(t_b,w_b).
\eeq
Thus, combining this with the expression for $\dot t_b$, one finds
that the brane trajectory in spacetime is determined by
integrating the first-order system
\beq
{dt_b\over d\tau}= {e^{-A_0}\over\sqrt{1-f^2}}, \qquad {d w_b\over
d\tau}={f\over\sqrt{1-f^2}}, \label{traj}
\eeq
with some initial conditions $t(0)=t_0$ and $w(0)=w_0$.
Finally the energy density on the brane is read off from the
isotropic junction condition,
\beq
{\kappa^2\over 6}\rho = e^{-A_0}\dot A {f\over \sqrt{1-f^2}} +
{A'\over \sqrt{1-f^2}}. \label{energy}
\eeq
Notice that $\rho$ is completely determined by the
{\it position} of the brane in the bulk spacetime. In order to get
its evolution as a function of $\tau$, one must solve for the
trajectory from (\ref{traj}) and then substitute into
(\ref{energy}).

As an illustration, let us consider the case $q_0=-{\sqrt{3}/4}$
which corresponds to a brane moving towards increasing values of
$w$ according to (\ref{wdot}) (we implicitly  assume that the bulk
is endowed with coordinates $t$ and $w$ that are positive).
Solving for the cosmological evolution on the brane, one finds
that as $\tau \rightarrow \infty$, $\rho$ converges towards the
Randall-Sundrum tension $\sigma_{\rm RS}=6/(\kappa^2\ell)$.  This
is shown by the solid line in figure 1 which plots the
``effective'' energy density $\rho_{\rm eff}=\rho-\sigma_{\rm RS}$
as a function of $\tau$. This can be understood by the fact that
the bulk geometry, at large $w$, resembles the warped Kasner
geometry (\ref{warped_Kasner}). We have also plotted the effective
pressure $P_{\rm eff}=P+\sigma_{\rm RS}$, which can be computed
from $\rho$ and the average Hubble parameter via the usual
energy conservation equation.  Observe that the effective pressure
is negative but converges towards zero more rapidly than the
effective energy, so that the effective equation of state
asymptotically corresponds to that of non-relativistic matter.
\begin{figure}
\begin{center}
\includegraphics[width=4.8in]{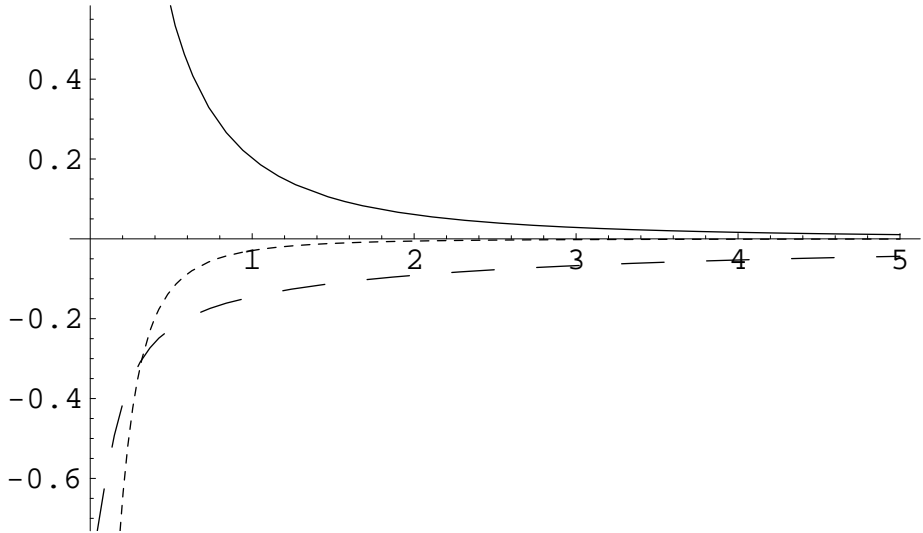}
\end{center}
\caption{The effective energy density $\rho_{\rm eff}$ (solid
line) and pressure $P_{\rm eff}$ (dashed line), in units of
$\sigma_{\rm RS}$, and the effective equation of state $P_{\rm
eff}/\rho_{\rm eff}$ (dotted line) as functions of the brane
cosmic time $\tau$. The plot corresponds to the initial conditions
$t_{\rm b}(0)=0.1$ and $w_{\rm b}(0)=0.2$.}
\end{figure}

\section{Conclusion}
In this paper we have constructed explicit anisotropic brane
cosmologies. We began our study by solving the full
five-dimensional vacuum  Einstein equations with negative
cosmological constant for a bulk metric admitting a homogeneous
but anisotropic (Bianchi I) three-dimensional slicing. Since the
general equations are too difficult, we have specialised our
analysis to two particular cases. The first case, in which the
bulk anisotropy was assumed to be only time-dependent, lead us to
bulk solutions that are {\it warped versions of 4D vacuum
solutions of Einstein equations with a cosmological constant}.
These solutions were already known. Then we assumed separability
of the metric components into time and extra-dimension dependent
pieces. In this way, we obtained, to our knowledge, {\it  new bulk
solutions that combine
 the 4D Kasner solution and the static 5D solutions of \cite{CMMS}}.

We then turned to the initial motivation for this work, namely
whether or not it is possible to embed an anisotropic brane with
only a perfect fluid as matter into a bulk geometry. Somewhat to
our surprise, we have found that it is possible to find such a
configuration in some of our ``hybrid'' bulk geometries: this is
because the anisotropic pressure on the brane induced by the bulk
time anisotropy can be compensated by a corresponding term induced
by the bulk spatial anisotropy when the brane is moving. This
compensation is possible only for particular trajectories, which
implies that there is no longer any flexibility in the choice of
the perfect fluid, in contrast with the isotropic case.  In some
sense, the perfect fluid on the brane is ``imposed'' by the bulk
geometry. In other words, it means that only a very particular
type of perfect fluid is compatible with the given bulk geometry.

This limitation is not surprising. It is simply the consequence of
having less symmetries. In the isotropic case, we have a high
level of symmetry which implies a generalized Birkhoff theorem:
because of the symmetries, the motion of the brane cannot perturb
the bulk geometry, in the same way as a moving spherical shell
cannot generate gravitational waves in 4D Einstein gravity. As
soon as we allow for anisotropy on the brane, its motion in the
bulk should  generate a very complicated bulk. In this respect, it
is a rather good surprise  that there exists an
 analytical bulk solution that allows for an anisotropic cosmology in
a purely perfect fluid brane.

\section{Acknowledgements}

We thank C.~Charmousis, R.~Durrer, S.~Foffa, R.~Maartens,  C.~Ringeval,
R.~Sturani and F.~Vernizzi for numerous useful discussions.  AF acknowledges
the University of Bologna for a fellowship under the program Marco Polo
at the Institut d'Astrophysique de Paris where this work was initiated. DAS
thanks Ruth Durrer for a wonderful conference in Sils Maria
during which many of these discussions took place.

\appendix
\section{Appendix}
Our purpose here is to give explicitly the solutions of the constraint
equations
\beq
 \sum_i p_i=1 , \quad \sum_i p_i^2=1, \quad
\sum_\mu q_\mu=0,  \quad \sum_\mu q_\mu^2={3\over 4},
\quad \sum_i q_i\left(p_i+1\right)=0.
\eeq
One can check that one can parametrize the solutions of these five
constraints as follows:
\beq
p_i={\sqrt{6}\over 3}\,  r_i(\phi)+{1\over 3},
\quad
q_i=\sqrt{{3\over 4}-{4\over 3}q_0^2}\,  r_i(\phi+\theta) -{q_0\over 3},
\eeq
with
\bea
r_1(\phi)&=&{\sqrt{3}+3\over 6}\cos\phi +{\sqrt{3}-3\over 6}\sin\phi, \\
r_2(\phi)&=&{\sqrt{3}-3\over 6}\cos\phi +{\sqrt{3}+3\over 6}\sin\phi, \\
r_3(\phi)&=&-{\sqrt{3}\over 3}\left(\cos\phi +\sin\phi\right),
\eea
and
\beq
\theta=\cos^{-1}\left({4\sqrt{2} q_0\over \sqrt{9-16 q_0^2}}\right).
\eeq
All allowed sets of coefficients are thus expressed in terms of two
parameters, $q_0$ and the angle $\phi$.

In the last section, we considered the particular case $q_0=\pm \sqrt{3}/4$.
Substituting in the above results, this yields the following parametrization:
\beq
p_i={\sqrt{6}\over 3}\,  r_i(\phi)+{1\over 3},
\quad
q_i={\sqrt{2}\over 2}\,  r_i(\phi+\theta_\pm) \mp {\sqrt{3}\over 12},
\eeq
with
\beq
\theta_+=0, \quad \theta_-=\pi.
\eeq

\section{Appendix}
In order to analyse the intrinsic properties of a metric, it is useful to
construct gauge-invariant quantities. A particular useful quantity in the
present context, where the Ricci tensor is already known, is the square
of the Weyl tensor.

For the metric  (\ref{metric_z}), we find that
the square of the bulk Weyl tensor is given by
\bea
&&C_{ABCD}C^{ABCD}= {1\over 3\ell^4}
\left\{
-{8\over 9}\ell^4
\left(-2+\sqrt{2}\cos(3\phi)-\sqrt{2}\sin(3\phi)\right)
 z^{-4q_0-1} (1-z^2) t^{-4}\right. \cr
&& \left.
+\left[ {16\over 9} q_0^2(-27+56 q_0^2)(1-z^2)^2
+8 q_0(27-80 q_0^2)(1-z^4)+{9\over 2}(23+2z^2+23 z^4)\right. \right.\cr
&& \left. \left.
-{2\over 9}\left(9-16 q_0^2\right)^{3/2}(1-z^2)\left(4 q_0(1-z^2)
+9 (1+z^2)\right)\left(\cos(3\phi+3\theta)- \sin(3\phi+3\theta)\right)\right]
z^{-4}(1-z^2)^2  \right. \cr
&& \left.
- {4\over 9}\ell^2 \left(9-16 q_0^2\right)(1-z^2)\left(\sqrt{2}
-\cos(3\phi)+\sin(3\phi)\right) z^{-5/2- 2q_0}(1-z^2)^{3/2}t^{-2}
\right\}.
\eea
Note that the spacetime becomes singular when $t \rightarrow 0$ except when
$3\phi+\pi/4=0$, which corresponds to one of the $p_i$ equal to $1$ while
the other two vanish.

\end{document}